\documentclass[a4paper,11pt]{article}
\usepackage{pos}
\usepackage{aas_macros}
\usepackage[outercaption]{sidecap}    
\usepackage{wrapfig}

\title{Survey of the Galactic Plane with the Cherenkov Telescope Array
}
 \ShortTitle{CTA-GPS simulations}

\author*[a]{Q.~Remy}
\author[b]{L.~Tibaldo}
\author[c]{F.~Acero}
\author[d]{M.~Fiori}
\author[b]{J.~Knödlseder}
\author[d]{B.~Olmi}
\author[f]{P.~Sharma}

\affiliation[a]{Max-Planck-Institut fur Kernphysik, Saupfercheckweg 1, 69117 Heidelberg, Germany}
\affiliation[b]{IRAP, Université de Toulouse, CNRS, UPS, CNES, F-31028 Toulouse, France}
\affiliation[c]{AIM, CEA, CNRS, Université Paris-Saclay, Université Paris Diderot, Sorbonne Paris Cité, F-91191 Gif-sur-Yvette, France}
\affiliation[d]{INAF - Osservatorio Astronomico di Padova, Vicolo dell’Osservatorio 5, I-35122, Padova, Italy}
\affiliation[e]{INAF - Osservatorio Astrofisico di Arcetri, Largo E. Fermi, 5 - 50125 Firenze, Italy}
\affiliation[f]{IJCLab, Université Paris-Saclay, Université de Paris, IN2P3/CNRS, 91405 Orsay, France}
\forColl{CTA Galactic Science Working Group} 

\emailAdd{quentin.remy@mpi-hd.mpg.de}
\emailAdd{luigi.tibaldo@irap.omp.eu}

\abstract{Observations with the current generation of very-high-energy gamma-ray telescopes have revealed an astonishing variety of particle accelerators in the Milky Way, such as supernova remnants, pulsar wind nebulae, and binary systems. The upcoming Cherenkov Telescope Array (CTA) will be the first instrument to enable a survey of the entire Galactic plane in the energy range from a few tens of GeV to 300 TeV with unprecedented sensitivity and improved angular resolution. In this contribution we will revisit the scientific motivations for the survey, proposed as a Key Science Project for CTA. We will highlight recent progress, including improved physically-motivated models for Galactic source populations and interstellar emission, advance on the optimization of the survey strategy, and the development of pipelines to derive source catalogues tested on simulated data. Based on this, we will provide a new forecast on the properties of the sources that CTA will detect and discuss the expected scientific return from the study of gamma-ray source populations.
}

\FullConference{37$^{\rm{th}}$ International Cosmic Ray Conference (ICRC 2021)\\
		July 12th -- 23rd, 2021\\
		Online -- Berlin, Germany}

\begin{document}
\maketitle

\section{Introduction}

In the last two decades, surveys providing larger samples of sources with lower selection bias have enabled major advances in the study of very-high-energy (VHE) gamma-ray sources \citep{2002A&A...395..803A, 2018A&A...612A...1H, 2018ApJ...861..134A, 2020ApJ...905...76A}.
A survey of the Galactic plane was proposed as a Key Science Project for CTA \citep{2013APh....43..317D,2019CTAscience} with the goals of providing an unprecedented census of VHE emitters in the entire Galactic plane, studying diffuse gamma-ray emission and searching for new and unexpected phenomena. This article presents recent progress made in the preparation of the project.

\section{Sky model}\label{sec:skymodel}

We have developed a new sky model comprised of three main components: a set of real sources modelled on the basis of observations from past and current instruments, synthetic populations for the three main classes of Galactic VHE gamma-ray emitters (PWNe, SNRs, and gamma-ray binaries) based on physical models informed by observations and the theory, and interstellar emission.

We model sources detected by existing IACTs based on the compilation provided by gamma-cat\footnote{\url{https://gamma-cat.readthedocs.io}}. We add sources detected by \textit{Fermi}~LAT based on the 3FHL catalogue \cite{2017ApJS..232...18A} and by HAWC based on the 2HWC catalogue \cite{2017ApJ...843...40A} not included in gamma-cat. For 14 of these sources the simple geometrical models (discs or Gaussians) used in the catalogues to describe their morphologies are replaced by more complex templates based on multiwavelength observations.
A sample of gamma-ray binaries and pulsars are modelled with dedicated temporal profiles.

Gamma-ray emission from young SNRs is modelled using a Monte~Carlo approach according to \cite{cristofari2017}. Emission from older SNRs interacting with the interstellar medium (ISNR) is modelled based on the same SNR population combined with a catalogue of observed Galactic molecular clouds \cite{rice2016}. 
PWNe are modelled based on the same SN progenitor population. For each Type~II SN with ejecta mass $\geq 5 M_\odot$ a pulsar is added with properties randomly drawn from the observed Galactic population of gamma-ray emitting pulsars \cite{watters2011} (most relevant for describing the young population of pulsars powering PWNe). The dynamics of the PWN/SNR systems are evolved over time through the free-expansion phase (using analytical models) and the reverberation and compression phases (using analytical approximations of hydrodynamical models). The spectral distribution of relativistic leptons and their inverse-Compton gamma-ray emission is evolved over time using the GAMERA library \cite{hahn2015} under the hypothesis of injection at a fixed fraction of the pulsar spin-down power. The PWN population model will be discussed in an upcoming publication \cite{fiori2021}.
The population of gamma-ray binaries (BIN) is modelled following \cite{2017A&A...608A..59D} to which we refer for details and references. The model is adjusted so that the progenitor distribution follows that used for the SNR and PWN models.

\begin{SCfigure}
    \includegraphics[width=0.59\textwidth]{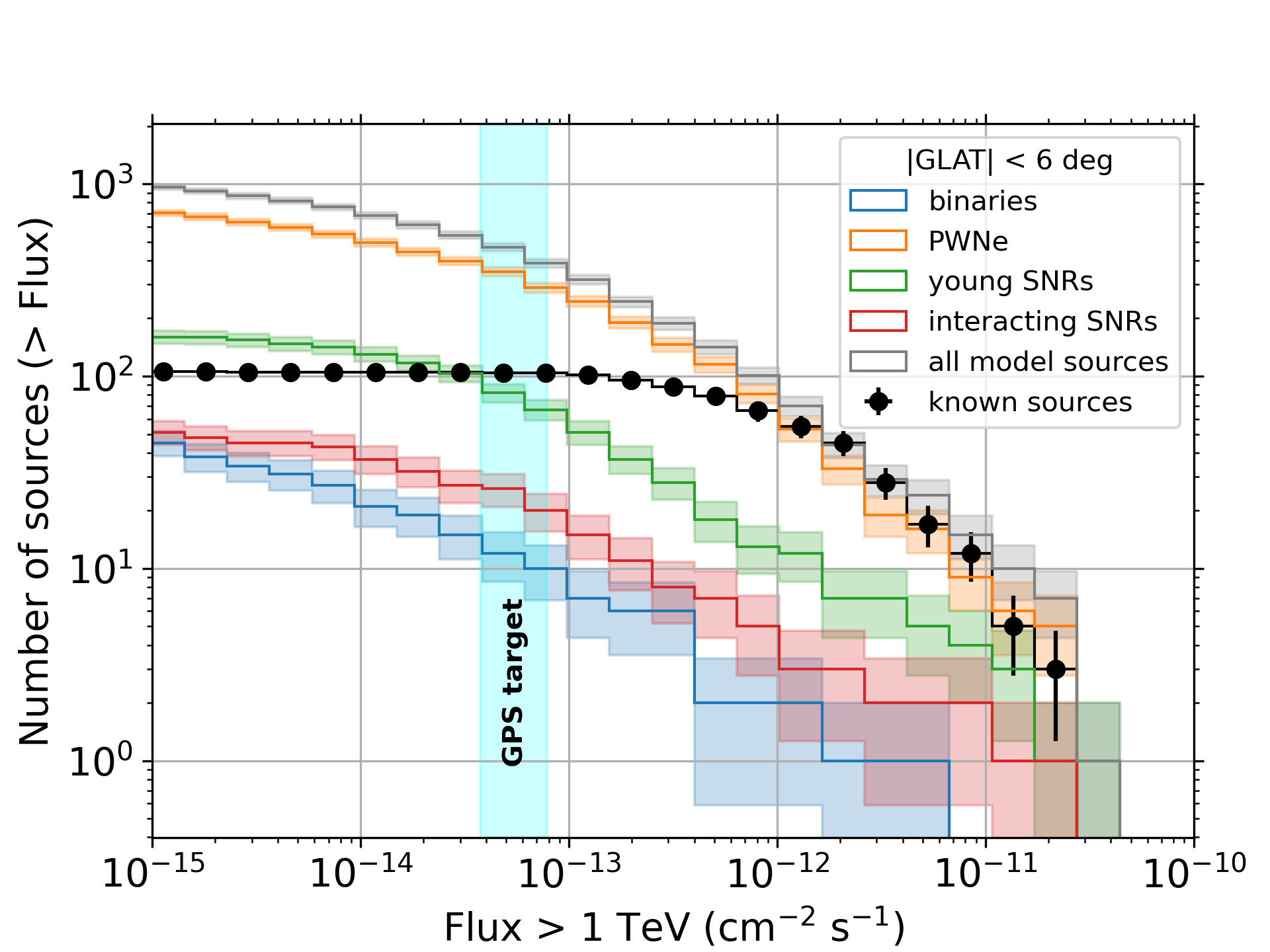}
    \caption{Cumulative source counts as a function of source photon flux integrated for energies above 1~TeV, as measured by existing and past IACTs (from gamma-cat), from \textit{Fermi}~LAT (3FHL), and HAWC (2HWC), and as predicted by the population synthesis for Galactic gamma-ray sources. Known or synthetic sources are only included if their distance from the Galactic equator is $< 6^\circ$. The vertical shaded band indicates the sensitivity target for the CTA GPS program \cite{2019CTAscience}.}        
    \label{fig:logNlogS_models}
\end{SCfigure}

Parameters of the four population models were tuned so that the bright end of the source count distribution as a function of flux reproduces the measured one for known sources, as illustrated in Figure~\ref{fig:logNlogS_models}. In this process it was assumed that sources currently unidentified are (predominantly) PWNe. The synthetic source populations include bright sources that could have already been detected with existing observations. We decide to retain the detected objects at the bright end of the flux distribution to have a more realistic sky model, therefore we need to remove some bright synthetic sources to preserve the overall source population properties. For each source already detected we exclude the most similar synthetic source belonging to the same class. The similarity is established based on source position, extension, and flux.

The interstellar emission model is based on the DRAGON cosmic-ray propagation code \cite{dragon2017} and computation of the related gamma-ray emission as described in \cite{hermes2021}. The model used here is tuned to direct cosmic-ray measurements near the Earth, but agnostic to current gamma-ray measurements which indicate larger/harder cosmic-ray spectra elsewhere in the Galaxy. It should therefore be considered as a minimal model for interstellar gamma-ray emission. Alternative models will be considered for our final study.

\subsection{Survey pointing strategy and simulations} 

\begin{figure}
    \centering
    \includegraphics[width=0.90\textwidth]{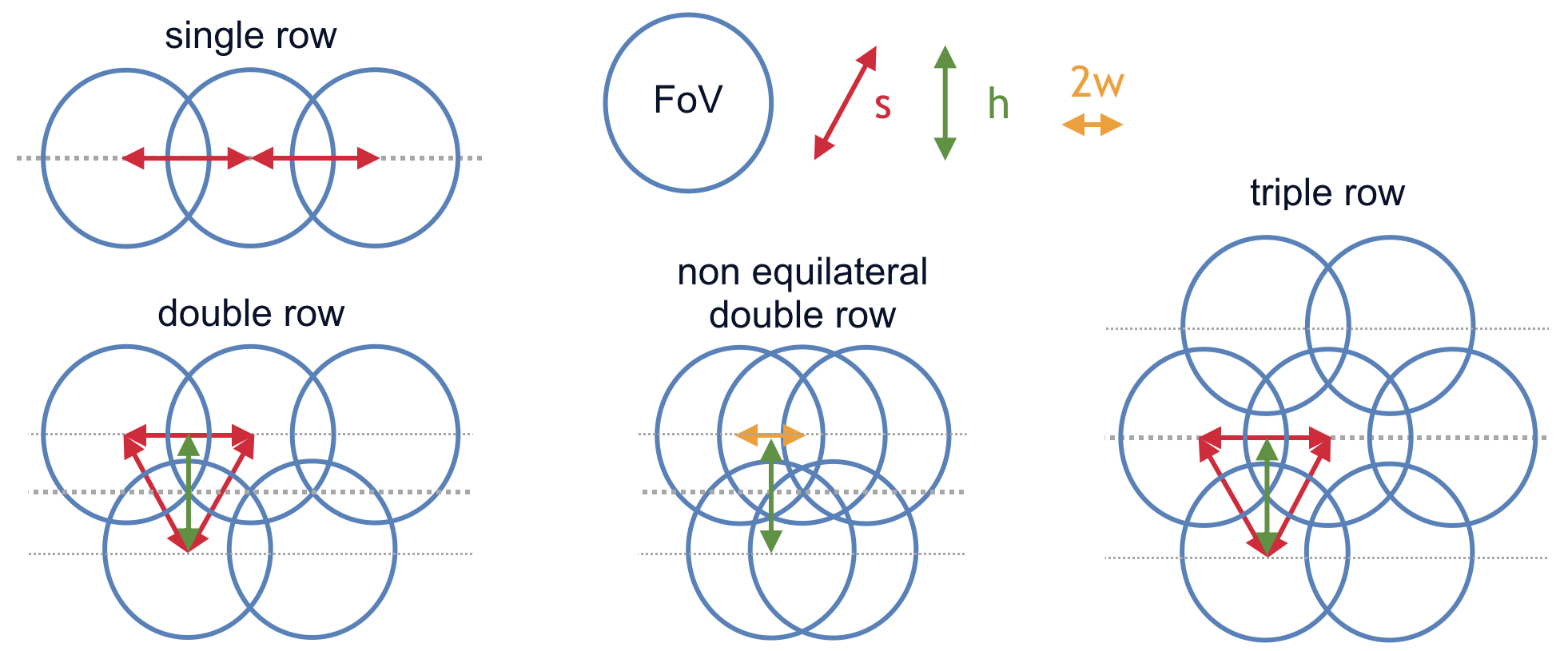}
    \caption{Schematic view of the pointing patterns considered. Dashed lines represent the Galactic equator. Circles represent the CTA field of view for an individual pointing. Red lines show the pattern step (s), i.e., the distance between pointing directions of adjacent pointings for single row and equilateral patterns. Green lines show the latitude spacing (h) for patterns with multiple rows. The orange line shows twice the longitude spacing (w) for the non equilateral double row pattern. For easier comparison with the other patterns we define the step of the non equilateral double row pattern as $\mathrm{s} = \sqrt{4/3} \, \mathrm{h}$, i.e., the step of an equilateral double row pattern with the same latitude spacing h.}    
    \label{fig:patterns}
\end{figure}

We explored a few different pointing patterns, including the single-row and double-row patterns from \cite{2013APh....43..317D,2019CTAscience}, as well as a triple-row pattern and a non equilateral double-row pattern with independent spacing in longitude and latitude. The pointing patterns are illustrated in Figure~\ref{fig:patterns}.
We compared for different pointing patterns and a few energy ranges the exposure uniformity, effective PSF, and sensitivity to a point-like isolated source using the {\em prod3b\_v2} IRFs.
We select for the rest of the study the non-equilateral double row pattern with a step $\mathrm{s} = 2.25^\circ$ ($\mathrm{h} = 1.95^\circ$) which provides the best sensitivity in the Galactic plane (with a broad minimum around the step value chosen) and almost as good sensitivity as the triple-row pattern at higher latitudes. The values of the longitude spacing w are set to cover the entire Galactic plane with the observing times partitioned as described in \cite{2019CTAscience} for individual observations lasting 30~minutes.

We implemented a realistic scheduling strategy following \cite{2019CTAscience} with a short-term programme during which 480 h of observing time were allocated over the first two years, and a long-term programme during which 1140 h of observing time were allocated over the following eight years. Observations were scheduled taking into account the Sun and Moon positions so that a given pointing is observed as close as possible to its minimum zenith angle. In addition, observations were distributed in time such that a given direction on the sky is revisited at different time intervals, enabling the detection of periodic source flux variations with periods between a few days and up to about 8 years.

Observations were simulated using the {\em ctools} package \citep{ctools} version 1.7.0. For the simulations, we used the {\em prod3b-v2} IRFs for the baseline CTA configuration optimized for observation durations of
50 hours for the zenith angle value closest to the actual zenith of each pointing.

\section{Catalogue}

\subsection{Analysis outlines}

\begin{figure}
    \centering
    \includegraphics[width=\textwidth]{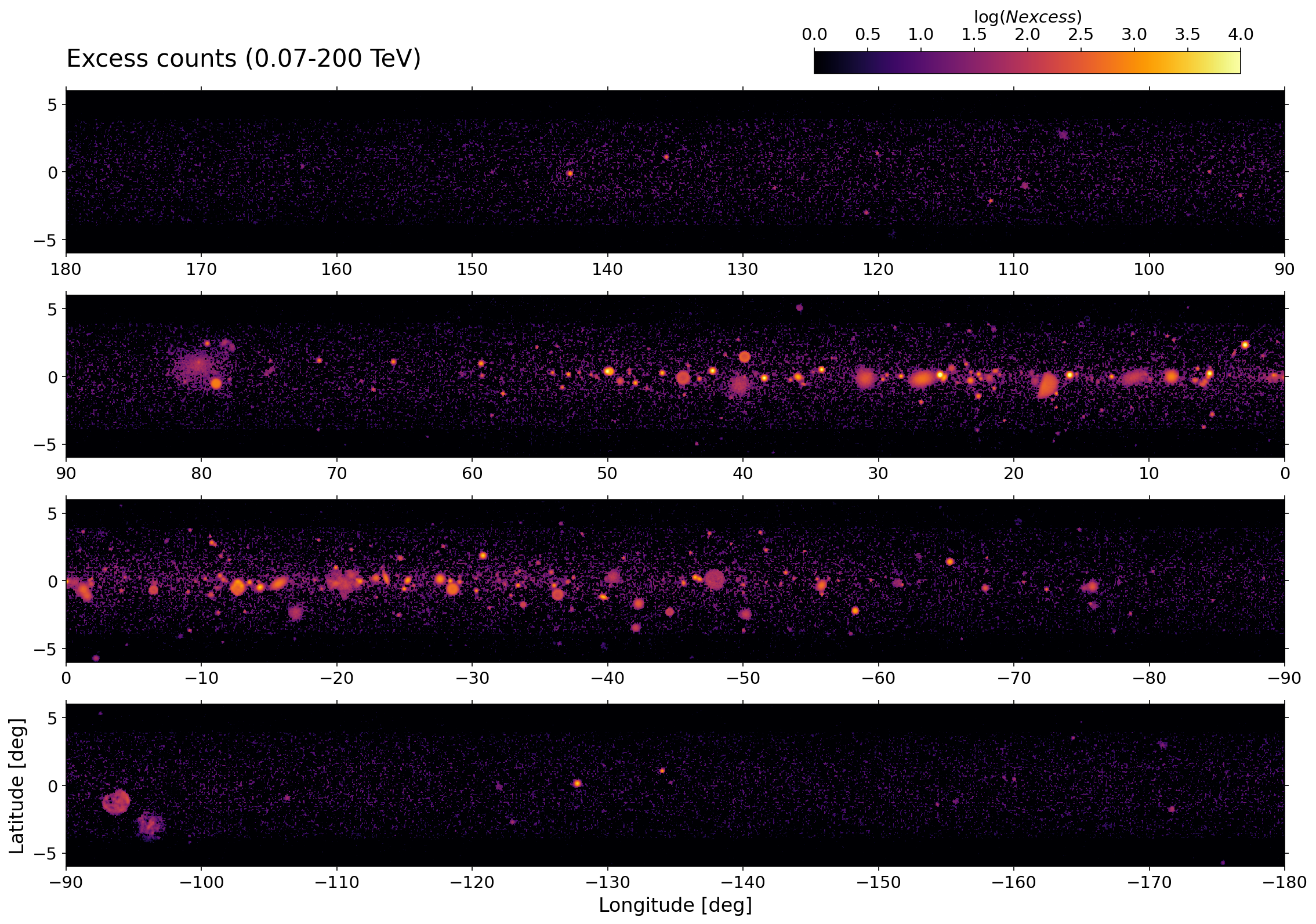}
    \caption{Excess counts above the instrumental background model in the 0.07-200 TeV energy range from the entire CTA GPS.}
    \label{fig:survey}
\end{figure}

Using the simulated data we built a catalogue of sources in the entire Galactic Plane at latitudes $|b|<6^\circ$
and for energies between 0.07 and 200~TeV. In Figure \ref{fig:survey} we show the excess counts above the instrumental background model for the full GPS survey in this energy range.

The first step of the catalogue production is to build in a short amount of computational time a list of candidate objects from the structures found in the excess or significance maps. The candidate objects are then fitted with different models in order to determine the best-fit model and its optimal parameters. The fitted candidates are filtered such that for each object $\rm{TS_{null}} = 2 \, \Delta ln(L) > 25$ with $\Delta ln(L)$ the difference in log-Likelihood between the best-fit model and the null hypothesis (no source).

The catalogue production was performed independently with {\em ctools} \citep{ctools} and  {\em Gammapy} \citep{gpy} analysis packages. In the following we will refer to the output catalogues as A and B, respectively. Note that the two analysis packages include the same features so the main differences between the output catalogues results from the analysis strategies used for the initial object detection (different algorithms), and the model fitting refinement (order and type of models tested). Catalogue A is based on the work presented in \cite{2018sf2a.conf..205C}. Catalogue B exploits the techniques and ideas discussed in \cite{2020APh...12202462R}, and extends beyond. In the following we will not discuss in detail the two strategies but in the next section we will comment on how the differences between them affect the results.


\subsection{Diagnostics and results}

In order to match the detected objects with the simulated sources we test for spatial coincidence using two criteria based on inter-center distance and surface overlap. For each object we search for simulated sources within an inter-distance $d_{c}<0.1+0.3\times R_{\rm object}$, and we report only the source maximizing the surface overlap fraction, defined as:
$\rm{SF_{\rm overlap}}= \left( S_{\rm object\,  \cap  \, source} \right) / \left( S_{\rm object \, \cup \, \rm source} \right)$,
where the surfaces are delimited by the iso-contours in flux of the PSF-convolved model corresponding to a 68\% containment.
We choose to report only the best association in order to limit the possible associations for extended objects. We also enforce that each source can be associated to only one object and vice-versa. Moreover, we report only the associations with $\rm SF_{ overlap}>0.25$ as this limits spurious associations. In order to test the overall quality of the catalogue produced we introduce the matching fraction,  $f_{\rm match}=N_{\rm match}/N_{\rm object}$, defined as the fraction of the detected objects matching a simulated source.

Table \ref{tab:detections} gives the numbers of simulated sources detectable with TS > 25 for the different synthetic source populations and for the known sources. Based on the matching criterion previously defined we also report the potential detections from the catalogues associated to the same populations and the overall matching fraction. These results show that we may detect up to 500 significant sources in the 0.07-200 TeV energy range from the CTA-GPS which is more than 6 times the number of objects in the HESS-GPS \cite{2018A&A...612A...1H} or the 3HAWC \cite{2020ApJ...905...76A} catalogues .

\renewcommand{\arraystretch}{1.}
\setlength{\tabcolsep}{0.18cm}
\begin{table*}
\caption{Number of detectable sources and detected objects with TS > 25 in the 0.07-200 TeV energy range.}
\centering
\begin{tabular}{c| c c c c c c c | c c}
\hline
\hline
Name &  PWN & SNR & ISNR & BIN & Known & No-match & Total & $f_{\rm match}$\\
\hline
Simulated detectable  & 294 & 37 & 24 & 10 & 134 & - & 499 & - & \\                                                         
Catalogue A  & 241 & 16 & 20 & 10 & 111 & 169 & 567  & 0.70 & \\
Catalogue B  & 257 & 31 & 14 & 10 & 122 & 36 & 470  & 0.92 & \\                      
\hline
\hline
\end{tabular}
\label{tab:detections}
\end{table*}

\begin{SCfigure}
    \includegraphics[width=0.62\textwidth]{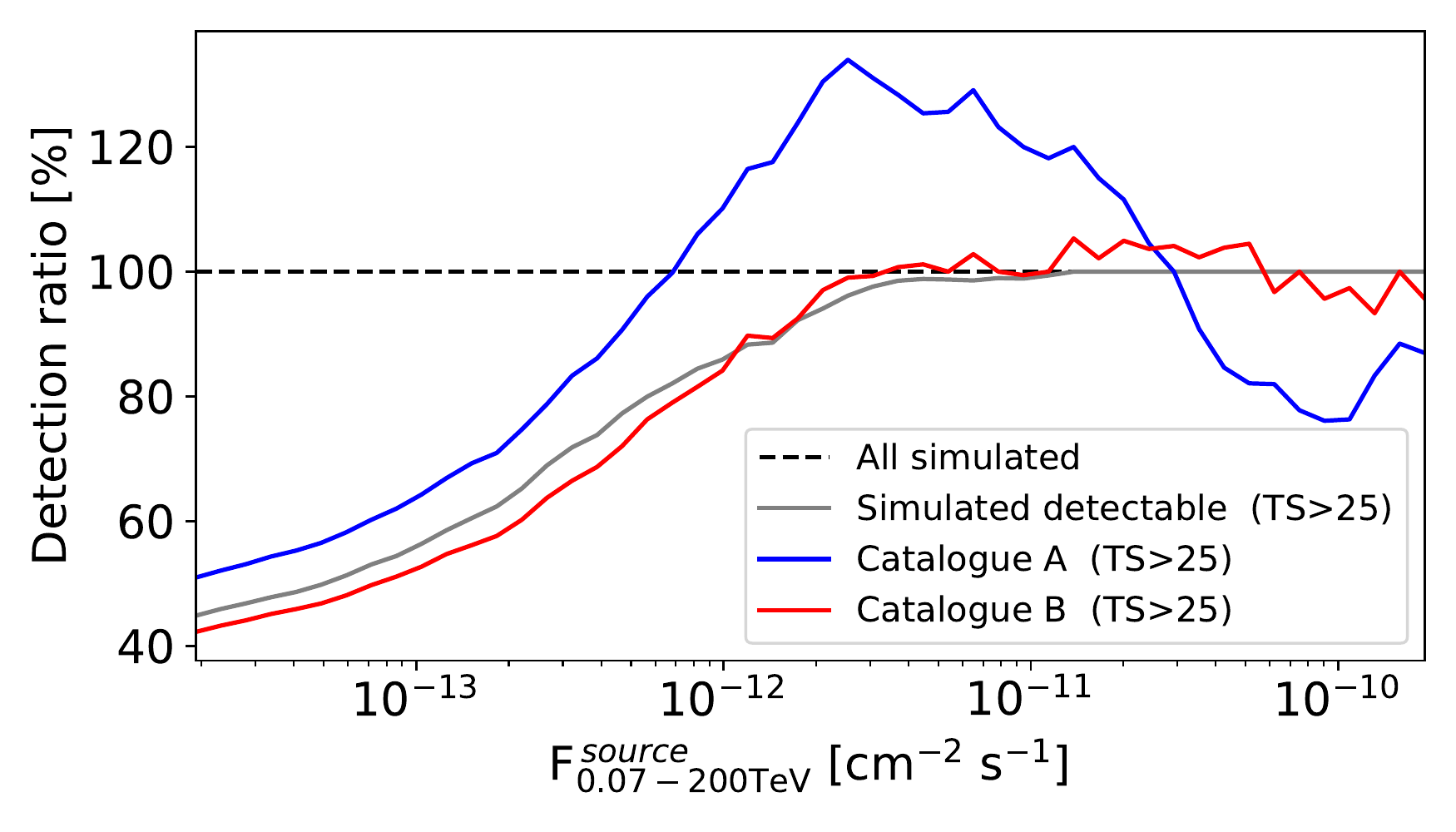}
    \caption{Detection ratio above a given integrated flux (for E = 0.07-200 TeV). The grey curve corresponds to the expected detections with TS > 25 for the exact simulated models, the colored curves correspond to the detections reported in the catalogues at the same threshold. Note that the detection ratio of a catalogue can exceed 100\% because of the confusion bias or modelling biases (see details in the text).}
    \label{fig:det_frac}
\end{SCfigure}

We also define the detection ratio as the number of objects detected with $\mathrm{TS} > 25$ divided by the total number of simulated sources. Figure \ref{fig:det_frac} shows the detection ratio above a given integrated flux as a function of flux. The detection ratio of a catalogue can exceed 100\% because of the confusion bias or modelling biases. In the case of the confusion bias, the emission from the sea of sources below the detection threshold biases upwards the flux of the sources near the threshold and so enhance their detection. In the case of the modelling bias, we see that sources simulated with more complex models (shell, elliptical Gaussian, or template) than the ones considered in the catalogue construction (Gaussian, disc, and point-like) can be fragmented into multiple smaller objects of lower flux. This fragmentation of the complex sources in multiple sub-structures explain mostly the discrepancy observed for the Catalogue A in Figure \ref{fig:det_frac}, and its larger number of objects that comes with a lower matching fraction as reported in \mbox{Table \ref{tab:detections}}. At this stage filtering and merging the spurious detections is more a classification problem than a statistical problem. Thus, the solution introduced in the production of Catalogue B to solve this issue is to use pattern recognition techniques : (i) to determine \textit{a priori} the most suited candidates to be fitted with more complex models such as shells or elliptical Gaussians; (ii) to identify  \textit{a posteriori} the groups of objects that could be merged together or replaced by a different model.
The results of Catalogue B in term of detection ratio and matching fraction show that we can produce a catalogue close to the absolute limit of detections expected from the exact simulated models.
 
\section{Source populations}

\begin{figure*}
    \centering
    \includegraphics[width=0.54\textwidth, bb= 0 0 420 320, clip]{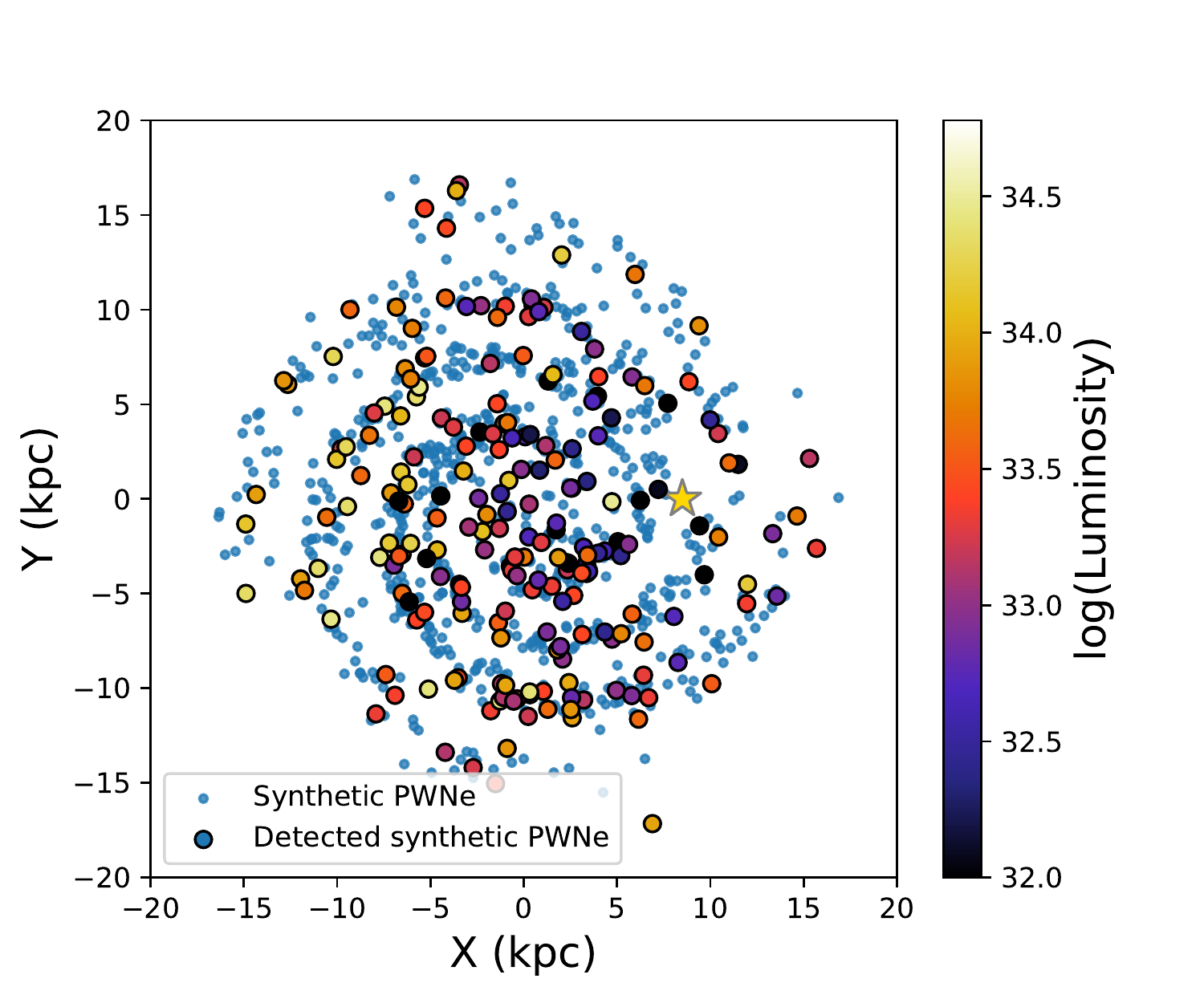}
    \includegraphics[width=.42\textwidth, bb= 0 0 325 318,clip]{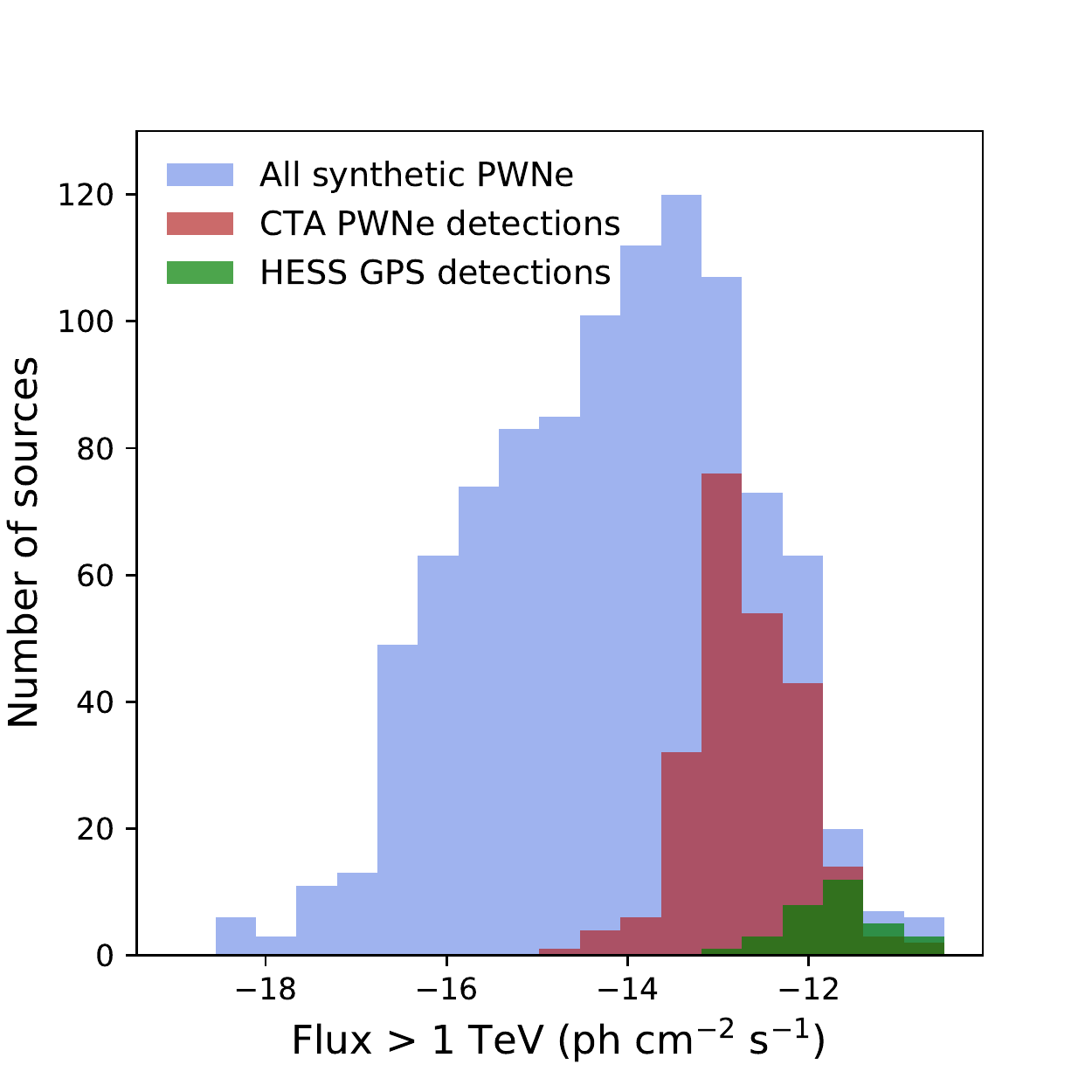}
        \vspace{-0.2cm}
    \caption{Properties of the synthetic PWN population. \textbf{Left panel}: Galactic spatial distribution of the synthetic and detected objects with a color coding in terms of luminosity ($>$ 1 TeV ; ph s$^{-1}$). Note that the void of detections in the solar neighborhood is due to nearby synthetic sources having similar properties to known sources being removed (see Sect.\ref{sec:skymodel}).  \textbf{Right panel}: Histogram of the CTA GPS detections compared to the entire synthetic PWN population and the HESS GPS PWN catalogue.}
    \label{fig:pop}
\end{figure*}

In this section we will discuss the properties of the two dominant source classes detected in the survey: PWNe and SNRs.
PWNe are the dominant source class at TeV energies in the Galactic plane. About 250 new PWN detections are reported in the catalogue described above and a brief overview of their properties is discussed below.

Figure \ref{fig:pop} (left panel) shows the spatial distribution of all the synthetic PWNe generated in our sky model along with the ones that were detected by the catalogue pipeline. The CTA GPS sensitivity makes it possible to detect a large number of objects even at the opposite edge of our Galaxy.
However, at the other side of the Galaxy a strong selection is observed and only the brightest objects are observed.

The comparison in Figure \ref{fig:pop} (right panel)
of our current population of known PWNe \citep[HESS PWN catalogue][]{2018A&A...612A...2H} with the CTA detected ones and the entire synthetic population emphasize the transformational jump in population size that CTA will bring to the field of PWN population studies (about 7 times the current sample, or 2.5 times if we consider that most of the unidentified sources are PWNe).
We note that in this exercise, the association of a detected source with its counterpart is done by cross-matching with the coordinates of sources in the Sky model (see Sect.\ref{sec:skymodel}).
In real conditions, association to PWNe will be noticeably more difficult as the underlying population of PWNe is unknown. Numbers listed here should therefore be viewed as an upper-limit of what can be achieved with perfect counterpart catalogues.

\begin{SCfigure*}
    \centering
    \includegraphics[width=0.525\textwidth]{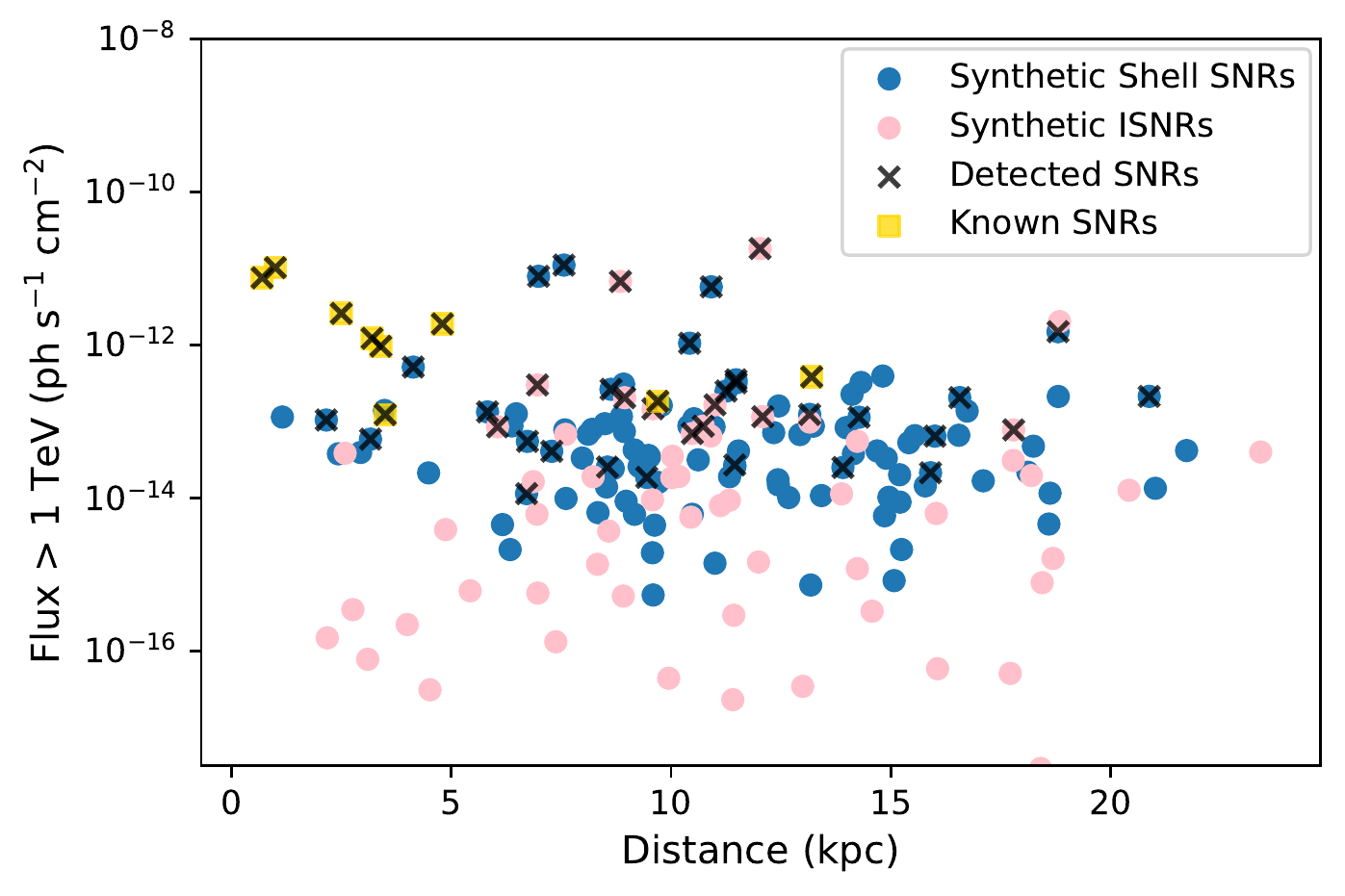}
    \caption{Comparison of the known SNRs and the population of synthetic shell and interacting SNRs in the Flux-distance parameter space. New objects can be detected up to a distance of 20 kpc and down to a flux of a few 10$^{-14}$ ph s$^{-1}$.}
    \label{fig:snr}
\end{SCfigure*}

The second most numerous class detected in this survey are SNRs. Focusing only on the synthetic shell SNRs and the synthetic interacting SNRs, 45 (31 and 14 respectively) sources have been detected in catalogue B. This suggests that the CTA GPS may be able to increase by a factor larger than two the SNRs observed at TeV energies. About half of the new detections are significantly extended which is a valuable feature for the identification of the newly discovered source with a multi-wavelength counterpart.
The distribution of SNRs in flux and distance in Figure \ref{fig:snr} shows that new sources can be detected up to the other side of our Galaxy and provide 5-10 times better flux sensitivity than the current TeV SNR sample.
This is a major step forward to explore the population of Galactic SNRs by discovering and measuring the extension of new SNRs even with the rather short exposure provided by the survey knowing that the current population of SNRs required deep $>100$~h exposures for the faintest objects.
\\

\small
\noindent \textbf{Acknowledgments.}  This work was conducted in the context of the CTA Galactic Science Working Group.
This research has made use of the CTA instrument response functions provided by the CTA Consortium and Observatory, see \url{http://www.cta-observatory.org/science/cta-performance/} (version prod3b-v2) for more details.
We gratefully acknowledge financial support from the agencies and organizations listed here:
\url{https://www.cta-observatory.org/consortium_acknowledgments/}.
The complete list of the CTA consortium members and their affiliations can be found here : \url{https://www.cta-observatory.org/consortium_authors/authors_2021_07.html}

\fontsize{8pt}{0pt}\selectfont

\bibliographystyle{aa}
\bibliography{references}



%
%
%

\end{document}